# Connecting bright and dark states through accidental degeneracy caused by lack of symmetry

Zixuan Hu[1,2], Gregory S. Engel[3], Sabre Kais*[1]

1. Department of Chemistry, Department of Physics, and Birck Nanotechnology Center, Purdue University, West Lafayette, IN 47907, United States
2. Qatar Environment and Energy Research Institute, College of Science and Engineering, HBKU, Qatar
3. Department of Chemistry, University of Chicago, Chicago, IL 60637, United States

*Email: kais@purdue.edu

Coupled excitonic structures are found in natural and artificial light harvesting systems where optical transitions link different excitation manifolds. In systems with symmetry, some optical transitions are allowed, while others are forbidden. Here we examine an excitonic ring structure and identify an accidental degeneracy between two categories of double-excitation eigenstates with distinct symmetries and optical transition properties. To understand the accidental degeneracy, a complete selection rule between two arbitrary excitation manifolds is derived with a physically motivated proof. Remarkably, symmetry analysis shows the lack of certain symmetry elements in the Hamiltonian is responsible for this degeneracy, which is unique to rings with size $N = 4l + 2$ ($l$ being an integer).

## I. Introduction

Symmetry of the Hamiltonian leads to degeneracy in its eigenspace. In simple chemical systems such as molecules, degeneracies can often be assigned to geometrical symmetries and characterized by symmetry groups of the molecular Hamiltonian. Other types of symmetries, e.g. time reversal symmetry and translational symmetry, can also produce degenerate eigenstates. However, not every degeneracy can be explained by simple symmetries of the Hamiltonian. Accidental degeneracy is commonly associated with some unidentified hidden symmetry of the system. Notable examples of accidental degeneracy include the bound state degeneracy in a hydrogen atom[1], the degeneracy in an infinite square potential well[2], and the Landau level degeneracy in cyclotron motion[3]. Accidental degeneracy, in addition to theoretical interest[4], has potential applications in material design[5, 6]. In this study, we examine an accidental degeneracy found in a coupled excitonic ring structure. Coupled excitonic structures are found in natural and artificial light harvesting systems, in which delocalized quantum states are extensively studied for their potential effect on energy transfer efficiency[7-22]. These quantum states have selection rules determined by symmetry which, for the single-excitation manifold, separate the optically active bright state from the optically inactive dark states. In a ring structure made of identical local two level systems, the boundary coupling condition is dependent on the parity of the number of excitations on the ring. Therefore, the double-excitation manifold is related to the single-excitation manifold in a unique way. Below, we show that the double-excitation manifold of an excitonic ring has an accidental degeneracy between two categories of eigenstates with distinct symmetries, such that the first category only couples to the bright state while the second category only couples to the dark state through optical transitions. These degenerate eigenstates may potentially mix without additional energy cost, producing hybrid eigenstates that optically connect the bright and dark states. Such an accidental degeneracy is unique to a ring with $N = 4l + 2$ ($l$ being an integer) sites. Analysis of the relationship between the geometry of the ring and the eigenstates reveals that the accidental degeneracy is due to the absence of certain symmetry elements in the Hamiltonian.

## II. Theory

In this study, we consider a ring structure composed of identical local exciton-supporting sites. Excitons are hardcore bosons that cannot occupy the same quantum state. The creation and annihilation operators $a_j^\dagger$ and $a_k$ of hardcore bosons observe the bosonic commutation relation $\left[a_j^\dagger, a_k\right] = 0$ when $j \neq k$, but when $j = k$, they observe the fermionic anticommutation relation $\left\{a_j^\dagger, a_j\right\} = 0$. Obviously, the Pauli raising and lowering operators satisfy these conditions if we identify $a_j^\dagger$ with $\sigma_j^+$ and $a_k$ with $\sigma_k^-$. The Hamiltonian of the ring is then:

$$H = \omega \sum_{i=1}^{N} \sigma_i^+ \sigma_i^- + S \sum_{i=1}^{N} (\sigma_i^+ \sigma_{i+1}^- + \sigma_{i+1}^+ \sigma_i^-) \tag{1}$$

where $\hbar = 1$, $\omega$ is the site energy, $S$ is the coupling strength, and $\sigma_{N+1}^\pm = \sigma_1^\pm$. Under Jordan-Wigner transformation, the Hamiltonian becomes:

$$H_{JW} = \omega \hat{n} + S \sum_{j=1}^{N-1} \left(c_j^\dagger c_{j+1} + c_{j+1}^\dagger c_j\right) - S\left(c_N^\dagger c_1 + c_1^\dagger c_N\right) e^{i\pi \hat{n}} \tag{2}$$

where $\hat{n}$ is the number operator for the total number of excitations on the ring, and $c_j^\dagger$ and $c_j$ are the transformed fermionic creation and annihilation operators, respectively. The solution to equation (2) is dependent on the parity of the excitation number due to the $e^{i\pi \hat{n}}$ term in the boundary condition. For a given excitation number n, we first find the creation operators for the single excitation component states (referred to as the component states thereafter):

$$\begin{aligned} C_k^+ &= \frac{1}{\sqrt{N}} \sum_{j=1}^{N} e^{i\frac{k\pi}{N} j} c_j^\dagger && \text{if } n \text{ is odd} \\ C_k^+ &= \frac{1}{\sqrt{N}} \sum_{j=1}^{N} e^{i\frac{(k+1)\pi}{N} j} c_j^\dagger && \text{if } n \text{ is even} \end{aligned} \tag{3}$$

where $k$ is an even number from 0 to $2N - 2$. We then construct the n-excitation eigenstates by selecting n number of $C_k^+$ operators to operate successively on the ground state:

$$\left|\psi_{k_1 k_2 \ldots k_n}\right\rangle = C_{k_1}^+ C_{k_2}^+ \ldots C_{k_n}^+ \left|0\right\rangle \qquad (4)$$

where $k_1 \neq k_2 \neq \ldots \neq k_n$. The energies of the n-excitation eigenstates are given by $E_{k_1 k_2 \ldots k_n} = \sum_{i=1}^{n}\left(\omega + 2S\cos\frac{k_i \pi}{N}\right)$ if n is odd and $E_{k_1 k_2 \ldots k_n} = \sum_{i=1}^{n}\left(\omega + 2S\cos\frac{(k_i+1)\pi}{N}\right)$ if n is even, where the total energy is the sum of the energies of the individual component states. For example, the single-excitation eigenstates are:

$$\left|\psi_k\right\rangle = \frac{1}{\sqrt{N}} \sum_{j=1}^{N} e^{i\frac{k\pi}{N}j} c_j^\dagger \left|0\right\rangle \qquad (5)$$

where each $\left|\psi_k\right\rangle$ has the energy $E(k) = \omega + 2S\cos\frac{k\pi}{N}$. The component states that are building blocks for the double-excitation states have the expression:

$$\left|\phi_k\right\rangle = \frac{1}{\sqrt{N}} \sum_{j=1}^{N} e^{i\frac{(k+1)\pi}{N}j} c_j^\dagger \left|0\right\rangle \qquad (6)$$

where the energy of an individual $\left|\phi_k\right\rangle$ is $\varepsilon(k) = \omega + 2S\cos\frac{(k+1)\pi}{N}$, and the double-excitation state formed by $\left|\phi_{k_1}\right\rangle$ and $\left|\phi_{k_2}\right\rangle$ has the energy $E(k_1, k_2) = \varepsilon(k_1) + \varepsilon(k_2)$. Without loss of generality, in the following discussion we set $\omega = 0$.

The single-excitation states can be classified by their symmetry group representations which determine their optical transition possibilities. Here, we categorize the single-excitation states according to the possibility that they will optically transition to the ground state via the optical coupling operator $J^+ = \sum_{i=1}^{N} \sigma_i^+$. In this way, we can easily connect the theoretical model to its practical application in excitonic systems. By simple algebraic evaluation of equation (5), only the $\left|\psi_0\right\rangle = \frac{1}{\sqrt{N}} \sum_{j=1}^{N} c_j^\dagger \left|0\right\rangle$ state has finite coupling to the ground through optical transitions: the optical transition dipole is $\Gamma_{kg} = \left|\left\langle\psi_k\right| \sum_{i=1}^{N} \sigma_i^+ \left|0\right\rangle\right|^2 = N\delta_{k0}$, where $\delta_{k0}$ is the Kronecker delta. $\left|\psi_0\right\rangle$ is therefore the bright state and all other $\left|\psi_k\right\rangle$ are dark states. Similarly, the double-excitation states can also be categorized according to how they optically connect to the single-excitation states. The first category of double-excitation states are those optically coupled to the bright state but not the dark states; the second category of double-excitation states are those optically coupled to the dark states but not the bright state. To identify each double-excitation state by its category, we invoke a complete selection rule between the n-excitation manifold and the (n+1)-excitation manifold:

If an n-excitation state is $|\psi_{t_1 t_2 ... t_n}\rangle = C^+_{t_1} C^+_{t_2} ... C^+_{t_n} |0\rangle$ where the component states $C^+_{t_i}$'s have quantum numbers $t_i$'s, and an (n+1)-excitation state is $|\psi_{s_1 s_2 ... s_{n+1}}\rangle = C^+_{s_1} C^+_{s_2} ... C^+_{s_{n+1}} |0\rangle$ where the component states $C^+_{s_i}$'s have quantum numbers $s_i$'s, then $|\psi_{t_1 t_2 ... t_n}\rangle$ is optically coupled to $|\psi_{s_1 s_2 ... s_{n+1}}\rangle$ if:

$$\sum_i t_i - \sum_i s_i = 2mN \qquad m = 0, \pm 1, \pm 2 ... \qquad (7)$$

Note that here the quantum numbers $t_i$'s and $s_i$'s are slightly different from the $k_i$'s in equations (3) and (4), in which each $k_i$ is always even but when it enters the exponent of $C^+_{k_i}$ it becomes $(k_i + 1)$ when n is even and remains $k_i$ when n is odd. In $C^+_{t_i}$ or $C^+_{s_i}$ each $t_i$ or $s_i$ enters the exponent as itself, but the parity of $t_i$ or $s_i$ can change dependent on the parity of n. If n is odd then $t_i$'s are even and $s_i$'s are odd; if n is even then $t_i$'s are odd and $s_i$'s are even.

Equation (7) can be extracted from pure algebraic calculations[23], but here we emphasize that the selection rule is a consequence of the rotational symmetry of the ring. For a ring consisting of N identical local sites with the same coupling strength, a physical equivalence exists among different sites because there is no fundamental difference between one site and another. In the Supplementary Information (SI) we prove with detail that the selection rule is indeed a phase matching condition imposed by the physical equivalence among local sites, thus giving equation (7) a physically motivated explanation.

Applying equation (7) to the single-excitation manifold and the double-excitation manifold, we identify each double-excitation state by its category on the energy ladder in Figure 1:

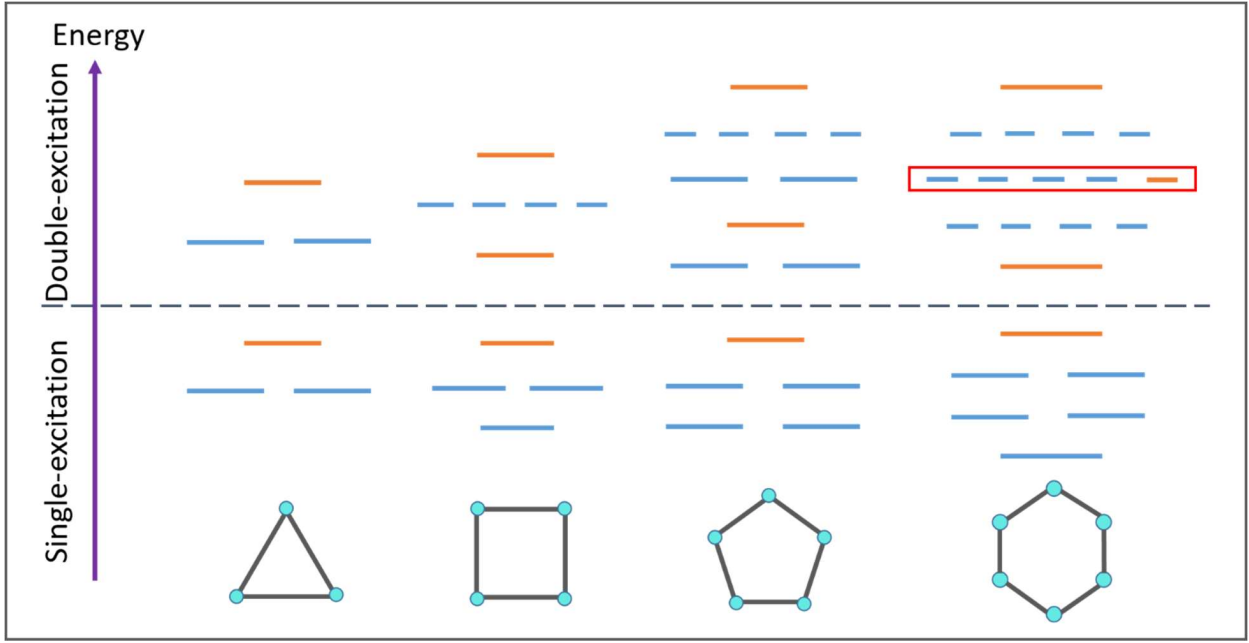

Figure 1. Energy ladder structure of the single-excitation and double-excitation eigenstates for rings with N = 3 to N = 6 sites. Single-excitation and double-excitation states are shown below and above the dashed line, respectively. The bright states and their optically connected double-

*excitation states are in orange (first category); the dark states and their optically connected double-excitation states are in blue (second category). The red box highlights the degeneracy between the two categories that is unique to the 6-sited ring.*

In Figure 1, we see the double-excitation manifold can be partitioned into two categories: 1) states optically coupled to the bright state and 2) states optically coupled to the dark states. For rings with 3 to 5 sites, the first category double-excitation states are non-degenerate, while the second category double-excitation states are evenly degenerate. For the 6-sited ring, however, there is a level on the double-excitation manifold with degeneracy five that contains states from both categories. This is an important property of the energy ladder because degeneracy may potentially allow states with different symmetries and optical transition properties to mix, producing hybrid eigenstates without additional energy cost. For example, if the degenerate states in the red box in Figure 1 are allowed to mix, the hybrid states would be able to mediate optical connection between the bright state and the dark states, which is otherwise forbidden due to symmetry mismatch.

To find out how the states from two categories are degenerate for the 6-sited ring, we consider how the double-excitation states are formed from the component states as defined in equation (6). The energy ladder structure of the component states is shown in Figure 2:

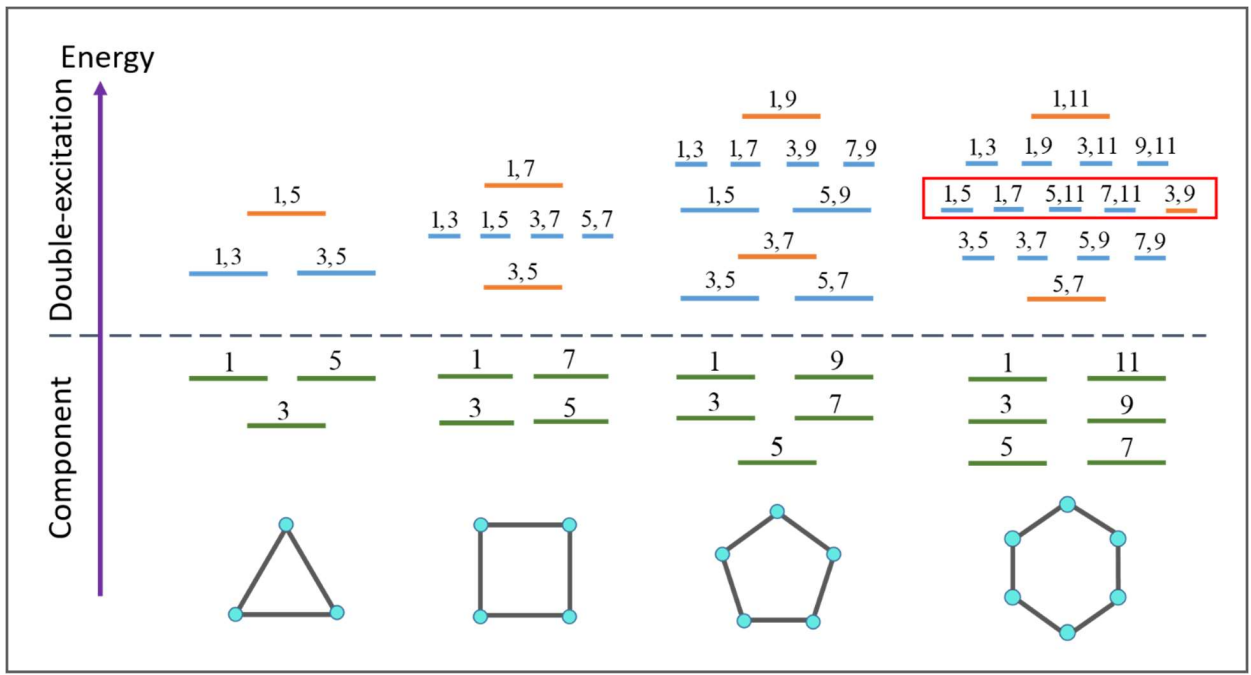

*Figure 2. Energy ladder structure showing how the double-excitation states are formed by the component states. Below the dashed line, the component states are labeled by single numbers corresponding to the $(k+1)$ values in equation (6). Above the dashed line, the double-excitation states are labeled by double numbers corresponding to their respective component states. The special level of degeneracy five is enclosed in the red box. Double-excitation states in the first category (orange) are formed by component states of the same energy; double-excitation states in the second category (blue) are formed by component states of different energies.*

Figure 2 shows clearly how the two categories of double-excitation states are separated. The first category states are formed by component states of the same energy and are non-degenerate for rings with 3 to 5 sites. These states can optically couple to the bright states, but not the dark states, because their symmetry only allows phase matching with the bright states (see the SI). On the other hand, the second category states are formed by component states of different energies and are evenly degenerate for rings with 3 to 5 sites. These states can optically couple to the dark states, but not the bright states, because their symmetry only allows phase matching with the dark states (see the SI). For the 6-sited ring, however, there is an accidental degeneracy between states of the two different categories which cannot be accounted for by simple symmetry of the Hamiltonian. Before we consider a symmetry argument responsible for the accidental degeneracy, notice that the accidental degeneracy happens because three energy levels in the 6-sited ring are evenly spaced in the component energy ladder, such that the sum of the energies of $|1\rangle$ and $|7\rangle$ is equal to the sum of the energies of $|3\rangle$ and $|9\rangle$. In the Supplementary Information, we prove with mathematical rigor that this scenario is possible if, and only if, the size of the ring $N$ satisfies

$$N = 4l + 2 \qquad l = 1, 2, 3, ... \qquad (8)$$

Therefore, the 6-sited ring is the first case of the accidental degeneracy - the next case would be the 10-sited ring. For a ring with an even number of sites, energy levels always equally split around zero, due to the sublattice symmetry of the Hamiltonian. Sublattice symmetry is immune to changes in the coupling strength $S$; therefore, the equal splitting behavior applies to both the single-excitation eigenstates and the component states, as can be seen from the energy diagrams below the dashed line in both Figure 1 and Figure 2. In Figure 2, for both the 4-sited ring and the 6-sited ring, the equally split pairs of component states combine to give zero energy double-excitation states of the second category. The unique accidental degeneracy is brought to the 6-sited ring by the zero energy component states, such as $|3\rangle$ and $|9\rangle$. These zero energy component states combine to give zero energy double-excitation states of the first category, which are only available when the size of the ring is $N = 4l + 2$. In Figure 3, for $N = 4, 6, 8$, we present both the single-excitation eigenstates and the component states in equations (5) and (6) on the complex number plane by representing each state with the factor in the coefficient: $e^{i\frac{k\pi}{N}}$ for the single-excitation eigenstates and $e^{i\frac{(k+1)\pi}{N}}$ for the component states:

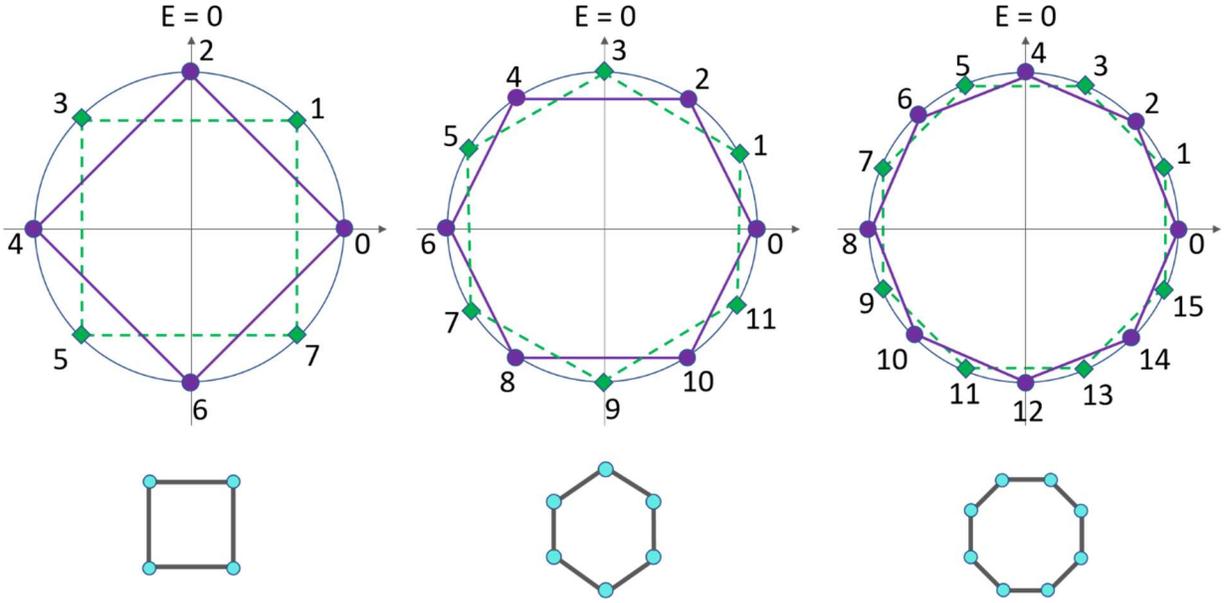

Figure 3. States are plotted on the complex plane. The $e^{i\frac{k\pi}{N}}$ points represent the single-excitation states (purple/circle labeled by even numbers). The $e^{i\frac{(k+1)\pi}{N}}$ points represent the component states (green/diamond labeled by odd numbers). The x-axis is the energy axis and the zero energy states fall on the y-axis. The single-excitation states form a shape corresponding to the physical geometry of the ring structures, and the component states form the same shape with a $\frac{\pi}{N}$ rotation. For the square and octagonal cases, zero energy states are in the single-excitation manifold and are missing from the component state manifold. For the hexagonal case, zero energy states are in the component state manifold and are missing from the single-excitation manifold.

In Figure 3, the single-excitation eigenstates form a shape corresponding to the physical geometry of the ring structures, while the component states form the same shape with a $\frac{\pi}{N}$ rotation. Because the energies of these states are calculated with the cosine function, degenerate states with energy $\varepsilon$ fall on the same vertical line of $x=\varepsilon$. The y-axis is, therefore, the zero energy line with $x=0$. Figure 3 shows that if there are zero energy states on the single-excitation manifold, there will be no zero energy component states. On the other hand, if there are zero energy component states, there will be no zero energy single-excitation state. The $\frac{\pi}{N}$ rotation between the single-excitation manifold and the component state manifold guarantees the zero energy degeneracy is only present in one of the two manifolds. When $N=4l$, the zero energy degeneracy is in the single-excitation manifold. When $N=4l+2$, the zero energy degeneracy is in the component state manifold. The accidental degeneracy between two categories of double-excitation states relies on the existence of degenerate zero energy component states for $N=4l+2$, which implies that there are no degenerate zero energy single-excitation states. There is a crucial difference between the single-excitation manifold and the component state manifold. The single-excitation states are actual eigenstates of the Hamiltonian, such that a degeneracy among them reflects a symmetry element of the Hamiltonian. On the contrary, the component states are not actual eigenstates of the Hamiltonian, and a degeneracy among them is not a symmetry element of the Hamiltonian. *Consequently the lack of zero energy degeneracy in the single-excitation manifold, which is the requirement for the accidental degeneracy, can be interpreted as the lack of certain symmetry in the Hamiltonian.* What exactly is the missing symmetry element in the $N = 4l + 2$ rings? In Figure 3, for all three geometries, there are two single-excitation states, $|\psi_{k=0}\rangle$ and $|\psi_{k=N}\rangle$, on the x-axis, which correspond to the $A_1$ and $B_1$ representations, respectively, in the $D_N$ symmetry groups ($D_{Nh}$ to be precise, but we do not mind the difference between gerade and ungerade here). For the $N = 4l$ rings, there are also two single-excitation states, $|\psi_{k=N/2}\rangle$ and $|\psi_{k=3N/2}\rangle$, on the y-axis, which result in a symmetry on the state diagram in Figure 3, such that if we rotate the shape formed by the single-excitation manifold by $\frac{\pi}{2}$, we get exactly the original shape back. Since the shape on the state diagram is the same as the physical geometry of the ring structures, we can relate this $\frac{\pi}{2}$ rotation to the $C_4$ symmetry elements in the $D_{N=4l}$ groups. Indeed, the $C_4$ symmetry elements are missing from the $D_{N=4l+2}$ groups. It is remarkable that the absence of certain symmetry elements, not the presence of one, leads to the accidental degeneracy between two categories of double-excitation states of distinct optical transition patterns.

To move from theory to application, the accidental degeneracy on the double-excitation manifold must be preserved under two types of disorder: the coupling strength disorder and the site energy disorder. Under disorder in the coupling strength $S$, the accidental degeneracy is exactly preserved for $N = 4l + 2$ with any arbitrary choice of the individual $S$ values. This remarkable preservation of the accidental degeneracy is guaranteed by the sublattice symmetry, which is intact as long as the ring topology of the Hamiltonian is maintained. When there is disorder in the site energy $\omega$, the perturbation to the Hamiltonian in equation (2) is $V = \sum_{j=1}^{N} \delta_j c_j^+ c_j^-$. The unperturbed component states for $N = 6$ are doubly degenerate, as shown in Figure 2; therefore, we use the degenerate perturbation theory to find the energy correction due to the perturbation. For all double-excitation states on the level of the accidental degeneracy, the first order energy correction is the same: $\varepsilon^{(1)} = \frac{2}{N} \sum_{j=1}^{N} \delta_j$, hence the accidental degeneracy is preserved up to first order site disorder (see the SI for details).

### III. Conclusion

In this work, we have identified and investigated an accidental degeneracy on the double-excitation manifold of a coupled excitonic ring structure. The accidental degeneracy occurs between two categories of double-excitation eigenstates which possess distinct symmetries and optical transition patterns to the single-excitation manifold. The accidental degeneracy has been proven to exist if, and only if, the size of the ring is $N = 4l + 2$. Using a state diagram relating the geometries formed by the single-excitation manifold and the component state manifold on the complex plane to the actual geometry of the ring, we have shown that, remarkably, the absence of certain symmetry elements is responsible for the accidental degeneracy. Finally the accidental degeneracy is immune to first order site energy disorder and arbitrary coupling strength disorder.

## IV. Acknowledgements

The authors thank Dr. Karen Watters for scientific editing of the manuscript. Funding for this research was provided by the Qatar National Research Foundation (QNRF), NPRP Exceptional Grant, NPRP X-107-010027.

# Supplementary information: connecting bright and dark states through accidental degeneracy caused by lack of symmetry

Zixuan Hu[1,2], Gregory S. Engel[3], Sabre Kais*[1]

4. *Department of Chemistry, Department of Physics, and Birck Nanotechnology Center, Purdue University, West Lafayette, IN 47907, United States*
5. *Qatar Environment and Energy Research Institute, College of Science and Engineering, HBKU, Qatar*
6. *Department of Chemistry, University of Chicago, Chicago, IL 60637, United States*

*Email: kais@purdue.edu

This supplementary document supports the discussion in the main text by providing technical details. Section 1 provides a physically motivated proof for the selection rule described by equation (7) in the main text. Section 2 gives a formal proof for the accidental degeneracy condition $N = 4l + 2$. Section 3 uses the degenerate perturbation theory to find the first order energy correction under site energy disorder.

1. **Physically motivated proof for the selection rule between different excitation manifolds**

We start with an analytical proof for the simple cases and then develop a physically motivated proof for the general case. For readers who want to skip the algebra, jump directly to the arguments after equation (1.10).

First consider the transitions from the single-excitation manifold to the double-excitation manifold. Single-excitation states have the form:

$$|\psi_k\rangle = \frac{1}{\sqrt{N}} \sum_{j=1}^{N} e^{i\frac{k\pi}{N}j} \sigma_j^+ |0\rangle \tag{1.1}$$

where $k$ is an even number from $0$ to $2N-2$. Double-excitation states have the form:

$$|\psi_{s_1 s_2}\rangle = C_{s_1}^+ C_{s_2}^+ |0\rangle = \frac{1}{N} \sum_{j<h}^{N} \left( e^{i\frac{\pi}{N}[s_1 j + s_2 h]} - e^{i\frac{\pi}{N}[s_1 h + s_2 j]} \right) \sigma_j^+ \sigma_h^+ |0\rangle \tag{1.2}$$

where $s_1$ and $s_2$ are odd numbers from $1$ to $2N-1$. The coupling between them is given by:

$$\begin{aligned}
\Gamma_{12} &= \left| \langle \psi_{s_1 s_2} | \sum_{j=1}^{N} \sigma_j^+ |\psi_k\rangle \right|^2 \\
&= \frac{1}{N^3} \left| \sum_{j<h}^{N} \left( e^{-i\frac{\pi}{N}[s_1 j + s_2 h]} - e^{-i\frac{\pi}{N}[s_1 h + s_2 j]} \right) \left( e^{i\frac{k\pi}{N}j} + e^{i\frac{k\pi}{N}h} \right) \right|^2 \\
&= \frac{1}{N^3} \left| \sum_{j<h}^{N} \left[ \left( e^{i\frac{\pi}{N}[(k-s_1)j - s_2 h]} - e^{i\frac{\pi}{N}[(k-s_1)h - s_2 j]} \right) + \left( e^{i\frac{\pi}{N}[-s_1 j + (k-s_2)h]} - e^{i\frac{\pi}{N}[-s_1 h + (k-s_2)j]} \right) \right] \right|^2
\end{aligned} \tag{1.3}$$

To evaluate (1.3), note the following equality:

$$\sum_{j<h}^{N} e^{i\frac{\pi}{N}(aj + bh)} = \sum_{m=1}^{N-1} \sum_{n=1}^{N-m} e^{i\frac{\pi}{N}[(a+b)m + bn]} = 2e^{i\frac{(a+b)\pi}{N}} \cdot \left(1 - e^{i\frac{a\pi}{N}}\right)^{-1} \cdot \left(1 - e^{i\frac{b\pi}{N}}\right)^{-1} \tag{1.4}$$

where both $a$ and $b$ are odd and $a + b \neq 2lN$ in which $l$ is an integer. Note equation (1.4) is invariant upon a-b switch. By making the connection $(k - s_1) + (-s_2) = (-s_1) + (k - s_2) = (a) + (b)$, we conclude equation (1.3) is always zero if $k - (s_1 + s_2) \neq 2lN$. When $a + b = 2lN$, equation (1.4) needs to be evaluated differently:

$$\sum_{j<h}^{N} e^{i\frac{\pi}{N}(aj + bh)} = \sum_{m=1}^{N-1} \sum_{n=1}^{N-m} e^{i\frac{\pi}{N}bn} = \left( Ne^{i\frac{b\pi}{N}} + 2 - N \right) \left( 2 - 2\cos\frac{b\pi}{N} \right)^{-1} \tag{1.5}$$

Using equation (1.5), after some algebra, equation (1.3) is then equal to:

$$\Gamma_{12} = \frac{1}{N} \left( \cot\frac{-s_2 \pi}{2N} + \cot\frac{(k - s_2)\pi}{2N} \right)^2 \neq 0 \tag{1.6}$$

where the inequality to zero is obtained by considering $k - (s_1 + s_2) = 2lN$ and $s_1 \neq s_2$. Therefore, we obtain a selection rule between the single-excitation manifold and the double-excitation manifold: for the transition to be possible, the quantum number $k$ from the single-excitation manifold minus the sum of the two quantum numbers from the double-excitation manifold must be equal to an integer multiple of $2N$. Can this selection rule be extended to higher excitation manifolds? Numerical results show that indeed for transitions between the double-excitation manifold and the triple-excitation manifold, the same selection rule holds: if the double-excitation state is made from two component states having quantum numbers $s_1$ and $s_2$, and the triple-excitation state is made from three single-excitation eigenstates having quantum numbers $k_1$, $k_2$ and $k_3$, then the transition is only possible when $(s_1 + s_2) - (k_1 + k_2 + k_3) = 2lN$. Considering the simplicity of the selection rule and its applicability to both the single-to-double and the double-to-triple transitions, we hypothesize that the general selection rule for transitions from any n-excitation manifold to the (n+1)-excitation manifold is:

$$\sum_i k_i - \sum_i s_i = 2lN \tag{1.7}$$

where the $k_i$'s and $s_i$'s are quantum numbers of the single-excitation states and component states that form the multi-excitation states in their respective manifold. To prove the general selection rule, consider an arbitrary triple-excitation state:

$$\left|\psi_{k_1 k_2 k_3}\right\rangle = C_{k_1}^+ C_{k_2}^+ C_{k_3}^+ \left|0\right\rangle$$

$$= \frac{1}{N^{3/2}} \sum_{j<h<n}^{N} \begin{pmatrix} e^{i\frac{\pi}{N}[k_1 j + k_2 h + k_3 n]} - e^{i\frac{\pi}{N}[k_1 h + k_2 j + k_3 n]} + e^{i\frac{\pi}{N}[k_1 h + k_2 n + k_3 j]} \\ -e^{i\frac{\pi}{N}[k_1 n + k_2 h + k_3 j]} - e^{i\frac{\pi}{N}[k_1 j + k_2 n + k_3 h]} + e^{i\frac{\pi}{N}[k_1 n + k_2 j + k_3 h]} \end{pmatrix} \sigma_j^+ \sigma_h^+ \sigma_n^+ \left|0\right\rangle \tag{1.8}$$

where the six terms in the sum are from the permutations of the three site indices $j$, $h$, and $n$, with the signs consistent with the fermionic behavior after the Jordan-Wigner transformation. The coupling of $\left|\psi_{k_1 k_2 k_3}\right\rangle$ to $\left|\psi_{s_1 s_2}\right\rangle$ is:

$$\Gamma_{23} = \left| \left\langle \psi_{k_1 k_2 k_3} \left| \sum_{j=1}^{N} \sigma_j^+ \right| \psi_{s_1 s_2} \right\rangle \right|^2 \tag{1.9}$$

in which

$$\sum_{j=1}^{N} \sigma_j^+ \left|\psi_{s_1 s_2}\right\rangle = \frac{1}{N} \sum_{j=1<h<n}^{N} \begin{pmatrix} e^{i\frac{\pi}{N}[s_1 j + s_2 h]} - e^{i\frac{\pi}{N}[s_1 h + s_2 j]} + e^{i\frac{\pi}{N}[s_1 h + s_2 n]} \\ -e^{i\frac{\pi}{N}[s_1 n + s_2 h]} + e^{i\frac{\pi}{N}[s_1 j + s_2 n]} - e^{i\frac{\pi}{N}[s_1 n + s_2 j]} \end{pmatrix} \sigma_j^+ \sigma_h^+ \sigma_n^+ \left|0\right\rangle \tag{1.10}$$

Due to the complexity of equations (1.8) and (1.10), direct evaluation of equation (1.9) in the same manner as equations (1.3) and (1.4) is very difficult. It would be case-specific and, therefore, not helpful for proving the general selection rule (1.7). In the following, we propose a physical argument to simplify the evaluation and generalize the proof.

Consider a ring consisting of N identical sites with the same coupling strength between sites. In such a model, there is a physical equivalence among different sites because there is no fundamental difference between one site and another: site $j$ behaves just the same way as site $h$. Considering the pair of sites $(j,h)$, it is clear that all the pairs with the same difference $d = h - j$ are physically equivalent. This formally implies that, for any eigenstate of the N-membered ring, the coefficients associated with equivalent multi-site basis vectors are equal up to a constant phase shift, which corresponds to a rotation of the ring. Consequently, we hypothesize that for the coupling strengths in equations (1.3) and (1.9) to vanish, the sums involving the coefficients of equivalent multi-site basis vectors must vanish separately. To verify the hypothesis, we go back to the transition between the single-excitation manifold and the double-excitation manifold and group the terms in equation (1.3) with the same $d = h - j$:

$$\Gamma_{12} = \frac{1}{N^3} \left| \sum_{j<h}^{N} \left( e^{-i\frac{\pi}{N}[s_1 j + s_2 h]} - e^{-i\frac{\pi}{N}[s_1 h + s_2 j]} \right) \left( e^{i\frac{k\pi}{N}j} + e^{i\frac{k\pi}{N}h} \right) \right|^2$$

$$= \frac{1}{N^3} \left| \sum_{d=1}^{M} \sum_{j=1}^{N} \left( e^{-i\frac{\pi}{N}[s_1 j + s_2 (j+d)]} - e^{-i\frac{\pi}{N}[s_1 (j+d) + s_2 j]} \right) \left( e^{i\frac{k\pi}{N}j} + e^{i\frac{k\pi}{N}(j+d)} \right) \right|^2 \quad (1.11)$$

$$= \frac{1}{N^3} \left| \sum_{d=1}^{M} \left[ \left( e^{-i\frac{\pi}{N}s_2 d} - e^{-i\frac{\pi}{N}s_1 d} \right) \left( 1 + e^{i\frac{\pi}{N}kd} \right) \right] \sum_{j=1}^{N} \left( e^{i\frac{\pi}{N}[j(k-s_1-s_2)]} \right) \right|^2$$

where $M = \frac{N}{2}$ or $M = \frac{N-1}{2}$, depending on the parity of $N$. Equation (1.11) has two separable sums and $\sum_{j=1}^{N} \left( e^{i\frac{\pi}{N}[j(k-s_1-s_2)]} \right)$ is zero unless $k - (s_1 + s_2) = 2lN$; therefore, we have recovered the selection rule proven previously. Now we can apply the same technique to equation (1.9) by grouping the terms with the same $d_1 = h - j$ and $d_2 = n - h$:

$$\Gamma_{23} = \left| \langle \psi_{k_1 k_2 k_3} | \sum_{j=1}^{N} \sigma_j^+ | \psi_{s_1 s_2} \rangle \right|^2$$

$$= \sum_{d_1} \sum_{d_2} f(d_1, d_2) \sum_j \left( e^{i\frac{\pi}{N}[j(s_1+s_2-k_1-k_2-k_3)]} \right) \quad (1.12)$$

where $f(d_1, d_2) = \left[ \begin{pmatrix} e^{i\frac{\pi}{N}[k_2 d_1 + k_3 d_2]} - e^{i\frac{\pi}{N}[k_1 d_1 + k_3 d_2]} + e^{i\frac{\pi}{N}[k_1 d_1 + k_2 d_2]} \\ -e^{i\frac{\pi}{N}[k_1 d_2 + k_2 d_1]} - e^{i\frac{\pi}{N}[k_2 d_2 + k_3 d_1]} + e^{i\frac{\pi}{N}[k_1 d_2 + k_3 d_1]} \end{pmatrix} \begin{pmatrix} e^{i\frac{\pi}{N}s_2 d_1} - e^{i\frac{\pi}{N}s_1 d_1} + e^{i\frac{\pi}{N}[s_1 d_1 + s_2 d_2]} \\ -e^{i\frac{\pi}{N}[s_1 d_2 + s_2 d_1]} + e^{i\frac{\pi}{N}s_2 d_2} - e^{i\frac{\pi}{N}s_1 d_2} \end{pmatrix} \right].$

Once again, equation (1.12) has the separable term $\sum_j \left( e^{i\frac{\pi}{N}[j(s_1+s_2-k_1-k_2-k_3)]} \right)$ equal to zero unless $(s_1 + s_2) - (k_1 + k_2 + k_3) = 2lN$. By equations (1.11) and (1.12), we see that selection rule (1.7) is indeed general because the physical equivalence among multi-site basis states remains true for any

13n-excitation manifold. In addition, the form $\sum_i k_i - \sum_i s_i = 2lN$ is a phase matching requirement for the initial and final states if any optical transition is possible.

## 2. Formal proof for the accidental degeneracy condition

In the main text, we have seen the accidental degeneracy can happen because, in the 6-sited ring, three energy levels are evenly spaced in the component state energy ladder, such that the sum of the energies of $|1\rangle$ and $|7\rangle$ is equal to the sum of the energies of $|3\rangle$ and $|9\rangle$. When does the component energy ladder have such a feature? First intuitively, the energy of a component state is given by $\varepsilon_m = \omega + 2S\cos\frac{(k+1)\pi}{N}$, and the energy ladder of the component states is symmetric with respect to $\frac{(k+1)\pi}{N} = \pi$. Without loss of generality, we focus only on the states with $\frac{(k+1)\pi}{N} \leq \pi$, and find this part of the energy ladder is antisymmetric with respect to $\frac{(k+1)\pi}{N} = \frac{\pi}{2}$. Therefore, by intuition we hypothesize that the three evenly spaced levels on the component energy ladder can exist if, and only if, there is a zero energy component state with $\frac{(k+1)\pi}{N} = \frac{\pi}{2}$, which is only possible when $N = 4l+2$. In the following, we provide a formal proof for this hypothesis. First, we abstract the statement into the following proposition:

> *Proposition: Consider the function $f(m) = \cos\left(\frac{2m+1}{N}\pi\right)$ where the integer $m$ satisfies $0 \leq m < N$ and the integer $N > 2$. The situation $f(m_1) - f(m_2) = f(m_2) - f(m_3)$ while $f(m_1) > f(m_2) > f(m_3)$ can happen if, and only if, $\frac{2m_2+1}{N} = \frac{1}{2}$ or $\frac{2m_2+1}{N} = \frac{3}{2}$ and $m_1 - m_2 = m_2 - m_3$.*

Here, we have replaced the even number $k$ with $2m$. Clearly, $f(m) = \cos\left(\frac{2m+1}{N}\pi\right)$ is the energy of the component states, and for the double-excitation states to have accidental degeneracy between the two categories, we must have three energy levels of the component states equally spaced as in $f(m_1) - f(m_2) = f(m_2) - f(m_3)$. The proposition outlined above provides a necessary and sufficient condition for this to happen. The $N = 4l+2$ rule for the accidental degeneracy then follows.

Proof: In Conway and Jones, Acta Arith. XXX (1976) 229-240, Theorem 7 (abbreviated to CJ7 in the following) states that:

*Suppose we have at most four distinct rational multiples of $\pi$ lying strictly between $0$ and $\pi/2$ for which some rational linear combination of their cosines is rational but no proper subset has this property. Then the appropriate linear combination is proportional to one from the following list:*

$\cos \pi/3 = 1/2$,
$-\cos \phi + \cos(\pi/3 - \phi) + \cos(\pi/3 + \phi) = 0$,
$\cos \pi/5 - \cos 2\pi/5 = 1/2$,
$\cos \pi/7 - \cos 2\pi/7 + \cos 3\pi/7 = 1/2$,
$\cos \pi/5 - \cos \pi/15 + \cos 4\pi/15 = 1/2$,
$-\cos 2\pi/5 + \cos 2\pi/15 - \cos 7\pi/15 = 1/2$,
$\cos \pi/7 + \cos 3\pi/7 - \cos \pi/21 + \cos 8\pi/21 = 1/2$,
$\cos \pi/7 - \cos 2\pi/7 + \cos 2\pi/21 - \cos 5\pi/21 = 1/2$,
$-\cos 2\pi/7 + \cos 3\pi/7 + \cos 4\pi/21 + \cos 10\pi/21 = 1/2$,
$-\cos \pi/15 + \cos 2\pi/15 + \cos 4\pi/15 - \cos 7\pi/15 = 1/2$.

In the current problem, we want to find the conditions for $\cos\left(\dfrac{2m_1+1}{N}\pi\right) + \cos\left(\dfrac{2m_3+1}{N}\pi\right) - 2\cos\left(\dfrac{2m_2+1}{N}\pi\right) = 0$, which for the moment can be written as $\cos\theta_1 + \cos\theta_3 - 2\cos\theta_2 = 0$, where $0 < \theta_i \leq \pi$. Before we can use CJ7, we need to consider the cases in which its assumptions are not satisfied. First, consider the case when $\theta_1 = \pi$ or $\theta_2 = \pi$. If $\theta_1 = \pi$, then $\cos\theta_3 - 2\cos\theta_2 = 1$, which is not possible by CJ7 if none of $\theta_2$ and $\theta_3$ is $\pi/2$. If indeed we allow $\theta_3 = \pi/2$ and $\theta_2 = 2\pi/3$, the equation is satisfied. However, in our original problem $\theta_i = \cos\left(\dfrac{2m_i+1}{N}\pi\right)$, therefore there are no integers of $N$ and $m_i$ that can make both $\theta_3 = \pi/2$ and $\theta_1 = \pi$, so this case is eliminated. On the other hand, if $\theta_2 = \pi$, then $\cos\theta_1 + \cos\theta_3 = -2$, which is impossible since we require $f(m_1) > f(m_2) > f(m_3)$.

Next, we consider the case where none of the three angles is $\pi$. Again, we need to first exclude the cases where one of the angles is $\pi/2$. Suppose $\theta_1 = \pi/2$, then $\cos\theta_3 - 2\cos\theta_2 = 0$, which is impossible by CJ7. On the other hand, if $\theta_2 = \pi/2$, then $\cos\theta_1 + \cos\theta_3 = 0$, which is impossible by CJ7 if $\theta_1$ and $\theta_3$ are distinct, as defined in CJ7. If $\theta_1$ and $\theta_3$ are equally spaced from $\pi/2$, $\theta_1 - \pi/2 = \pi/2 - \theta_3$, they will be considered not distinct by CJ7 and $\cos\theta_1 + \cos\theta_3 = 0$. Note that this case is just the one we originally proposed in the proposition: $\dfrac{2m_2+1}{N} = \dfrac{1}{2}$ and $m_1 - m_2 = m_2 - m_3$.

Finally, suppose none of the three angles are either $\pi$ or $\pi/2$. Then, if they are all distinct as defined by CJ7, we know $\cos\theta_1 + \cos\theta_3 - 2\cos\theta_2$ cannot be rational, and, therefore, cannot be zero. If two of the three angles are not distinct because they are equidistant from $\pi/2$, then either $-2\cos\theta_2 = 0$ or $\cos\theta_1 + 3\cos\theta_3 = 0$, which is impossible by CJ7.

To summarize, for all the cases where CJ7 does not apply, we have shown that the equality $\cos\left(\frac{2m_1+1}{N}\pi\right) + \cos\left(\frac{2m_3+1}{N}\pi\right) - 2\cos\left(\frac{2m_2+1}{N}\pi\right) = 0$ is satisfied if, and only if, $\theta_2 = \pi/2$ and $\theta_1 - \pi/2 = \pi/2 - \theta_3$. For the cases where CJ7 applies, we have shown that the equality is never possible. We conclude that the original proposition is true, giving the accidental degeneracy condition as $N = 4l + 2$, QED.

### 3. Degenerate perturbation theory treatment for site energy disorder

When there is disorder in the site energy $\omega$, the perturbation to the Hamiltonian in equation (2) in the main text is $V = \sum_{j=1}^{N} \delta_j c_j^+ c_j^-$. The unperturbed component states for the double-excitation states of the second category are a pair of $|C_{k_1}\rangle$ and $|C_{k_2}\rangle$ of opposite energies. Without loss of generality, we consider the pair $k_1 = 0$ and $k_2 = N-2$. Both $|C_{k_1}\rangle$ and $|C_{k_2}\rangle$ have their own degenerate counterparts and, to find the first order perturbation energy, we need to use the degenerate perturbation theory and diagonalize $P_{k_i} V P_{k_i}$, where $P_{k_i}$ is the projector into the degenerate space of $|C_{k_i}\rangle$. The result is a first order correction of $\alpha \pm \sqrt{|\beta|^2}$ for both the $|C_{k_1}\rangle$ and the $|C_{k_2}\rangle$ degenerate spaces, where $\alpha = \frac{1}{N}\sum_{j=1}^{N}\delta_j$ and $\beta = \frac{1}{N}\sum_{j=1}^{N}\delta_j e^{i\frac{2\pi j}{N}}$. Hence, by picking the eigenstate of energy $\varepsilon_{k_1} + \alpha + \sqrt{|\beta|^2}$ from the $|C_{k_1}\rangle$ degenerate space and the eigenstate of energy $\varepsilon_{k_2} + \alpha - \sqrt{|\beta|^2}$ from the $|C_{k_2}\rangle$ degenerate space, we can form a double-excitation state of the second category with the first order energy correction $E_I^{(1)} = \varepsilon_{k_1} + \varepsilon_{k_2} + 2\alpha$. The unperturbed component states for the double-excitation states of the first category are a pair of $|C_{k_3}\rangle$ and $|C_{k_4}\rangle$ of the same energy with $\frac{(k_3+1)\pi}{N} = \frac{\pi}{2}$ and $\frac{(k_4+1)\pi}{N} = \frac{3\pi}{2}$. Again, we use the perturbation theory for degenerate states and diagonalize $P_{k_3} V P_{k_3}$. The result is an energy shift of $\alpha \pm \gamma$ where $\gamma = \frac{1}{N}\sum_{j=1}^{N}\delta_j(-1)^j$, therefore, the double-excitation state formed by the eigenstates of $P_{k_3} V P_{k_3}$ has the first order energy correction of $E_{II}^{(1)} = 2\varepsilon_{k_3} + 2\alpha$. Because $2\varepsilon_{k_3} = \varepsilon_{k_1} + \varepsilon_{k_2}$, $E_{I}^{(1)} = E_{II}^{(1)}$, indeed the accidental degeneracy is preserved up to first order.